\documentclass[
    ,final            % use final for the camera ready runs
 ,numberedheadings % uncomment this option for numbered sections
  ]
  {aipproc}

\layoutstyle{8x11single}

\begin{document}

\title{Common Axioms for Inferring Classical Ensemble Dynamics and Quantum Theory}



\author{Rajesh R. Parwani\footnote{parwani@nus.edu.sg}}{
  address={Department of Physics and University Scholars Programme, National University of Singapore, Kent Ridge, Singapore.}
}

\begin{abstract}
 The same set of physically motivated axioms can be used to construct both the classical ensemble Hamilton-Jacobi equation and Schr\"odingers equation. Crucial roles are played by the assumptions of universality and simplicity (Occam's Razor) which restrict the number and type of of arbitrary constants that appear in the equations of motion.
In this approach, non-relativistic quantum theory is seen as the unique single parameter extension of the classical ensemble dynamics. The method is contrasted with other related constructions in the literature and some consequences of relaxing the axioms are also discussed: for example, the appearance of nonlinear higher-derivative corrections possibly related to gravity and spacetime fluctuations. Finally, some open research problems within this approach are highlighted.
 
\end{abstract}

\maketitle

%%%%%%%%%%%%%%%%%%%%%%%%%%%%%%%%%%%%%%%%%%%%
%% MAINMATTER
%%%%%%%%%%%%%%%%%%%%%%%%%%%%%%%%%%%%%%%%%%%%

\section{Introduction}

Over the years there have been several attempts to understand the foundations of quantum theory in more physical terms, or to re-formulate its mathematical structure. The various approaches have adopted perspectives from stochastic dynamics, statistical physics, probability theory, geometry, quantum information, and some more exotic disciplines. Recent examples of such endeavours may be found in various conference proceedings, for example \cite{qtrf1}, and in the online physics preprint archive. 

Some of the earliest attempts noted the close resemblance between  Schr\"odinger's equation, when written in some new variables, with other classical equations. After a transformation $\psi = \sqrt{p} \ e^{iS / \hbar}$, Schr\"odinger's equation reduces to the two coupled equations, 
\begin{eqnarray}
\dot{S} + {g_{ij} \over 2} \partial_iS \partial_j S + V + Q  &=& 0 \, ,  \label{hj3} \\
\dot{p}  + g_{ij} \ \partial_i \left( p \partial_j S \right)  &=& 0 \, ,  \label{cont} 
\end{eqnarray}
with 
\begin{equation}
Q \equiv -{{\hbar}^2 \over 8} g_{ij} \left( {2 \partial_i \partial_j p \over p} - {\partial_i p \partial_j p \over p^2} \right) \, 
\end{equation}
usually referred to as the quantum potential \cite{BH}. 

The notation is as follows: the equations have  been written in Cartesian coordinates for $N$ particles in $d+1$ dimensions, $i,j= 1,2,......,dN$ where $i=1,...d$, refer to the coordinates of the first particle of mass $m_1$, $i=d+1,.....2d$, to those of the second particle of mass $m_2$ and so on. The configuration space metric is diagonal and positive definite, $g_{ij} = \delta_{ij} /m_{(i)}$, with the symbol $(i)$ defined as   the smallest integer $\ge i/d$.  The overdot refers to a partial time derivative and the summation convention is used unless otherwise stated. 

Without the quantum potential, the first equation is recognizable as the Hamilton-Jacobi equation describing the classical motion of the particles, with $S$ determining the velocity, $v_i$,  of the particles through $v_i = g_{ij} \partial_j S$. 
Let us assume \cite{HR1} that one is uncertain about the initial conditions so that statistical methods must be adopted to locate the particles: $p(x,t)$ denotes the normalised probability density for the $N$ particles. Then the second equation of motion above is the continuity equation. Lest there be confusion, let me emphasize  that even for a single particle, $N=1$, one has an ensemble dynamics. However it is useful to discuss the $N>1$ case so that the separability axiom can be adopted later.

The pleasing picture of classical ensemble dynamics is blurred when the peculiar looking quantum potential is included. There have been numerous investigations to obtain and understand the structure of the quantum potential using some form of micro-dynamics, such as the stochastic approach of Ref.\cite{Nelson}; for more references see, for example, \cite{K}. However, as even the proponents of those ideas admit, some of the assumptions made of the micro-physics go beyond familiar classical dynamics; thus the mystery is not resolved but the focus  shifted to another area. 

An alternative, and perhaps less ambitious, route has been to understand the form of $Q$ without assuming any specific micro-dynamics, but rather using consistency conditions that are physically motivated: 
I will refer to this approach as ``physical axiomatics". In Sect.(2) I describe an example, from another branch of physics, on the use of physical axiomatics for constucting dynamical equations.
Following that, in Sect.(3), I review two recent attempts to apply an axiomatic approach to understand the form of $Q$. That sets the stage for a complete axiomatic construction of Schr\"odinger's equation, obtained in Ref.\cite{P5}, and summarised in Sect.(4). Some new topics are discussed in Sect.(5) while the concluding section lists some open problems for future investigation.

\section{An Example of Physical Axiomatics}
Here is a question that can be answered by using a constructive approach based on physically motivated axioms:
``What is the lagrangian for quantum electrodynamics (QED)?" 

The discussion can start with the electron and one can obtain for it, following Dirac, a relativistic equation which satisfies desired conditions. The Dirac equation has global gauge-invariance. Arguing for local-invariance, as Weyl and Yang-Mills did, then introduces a gauge-potential through a covariant derivative. Next, requiring the gauge field to be dynamical means that the lagrangian must have terms containing derivatives of the gauge-field. Imposing physical constraints such as locality, local gauge-invariance, Poincare invariance  and renormalisability, gives one the final $F_{\mu \nu}^{2}$ form quoted in textbooks. 

Of course the construction described above was not the historical route that led to QED but, once the various ingredients were recognised, similar arguments were used  later to construct the standard model of particle physics, and to propose various extensions probing physics beyond the standard model.

One may adopt a similar philosophy, of using physically motivated axioms, to try and understand the structure of {\it quantum theory itself} which is currently taken for granted in the standard model and its extensions. This is discussed in the next two sections.

\section{Deriving the Quantum Potential from Axioms} 

The derivations of the quantum potential in Refs.\cite{HR1,P1,Reg}  both  have as their common starting point the classical  ensemble action 
 \begin{eqnarray}
\Phi_C &=& \int  p \left[ \dot{S}  + {g_{ij} \over 2} \partial_i S \partial_j S   + V \right] dx^{Nd}  dt  \,. 
\label{classA} 
\end{eqnarray}
Varying the action (\ref{classA}) with respect to the field variables $p(x,t)$ and $S(x,t)$ (with $x$ summarising all the spatial coordinates) gives rise to 
\begin{eqnarray}
\dot{S} + {g_{ij} \over 2} \partial_iS \partial_j S + V &=& 0 \, ,  \label{hjc} \\
\end{eqnarray}
and the continuity equation (\ref{cont}).

The next step in both of the derivations \cite{HR1,P1} is to argue for an extension of (\ref{classA}) and  then to constrain the extension with some axioms so that a unique result is obtained. In this way a quantum action  results which gives rise to the quantum Hamilton-Jacobi equation and the continuity equation, and hence equivalently the  Schr\"odinger's equation. Planck's constant appears in these approaches to match the dimensions of the new term in the action with the old terms. 

Thus in both of the approaches \cite{HR1,P1}, essentially only the quantum potential is constructed from various axioms. However it was  shown later in Ref.\cite{P5} that one may use  a single set of intuitive axioms to {\it simultaneously} derive the classical equations and their quantum extensions which lead to Schr\"odinger equation. Thus both the classical ensemble theory and quantum theory result from the same set of axioms with quantum theory being a single parameter extension of the classical dynamics. One advantage of the unified approach of Ref.\cite{P5} is that it avoids the usual but peculiar textbook procedure  of obtaining quantum theory by  ``quantising" the  classical limit of that same theory, when clearly the latter  should be derived from the former. The method of Ref.\cite{P5} is summarised in the next section.

\section{The Schr\"odinger Equation from Scratch}

Suppose there are $N$ particles whose dynamics we would like to describe. It will be assumed that a configuration-space metric as described earlier exists. As before, suppose that we do not have sufficient information about the micro-dynamics to fix the initial conditions, so that statistical methods must be used to locate the particles. Indeed, the underlying dynamics could very well be deterministic but chaotic so that the trajectories become sensitive to initial conditions \cite{Stro} of which we lack knowledge of. For some recent papers discussing the emergence of quantum dynamics within various deterministic theories, see \cite{chaos}.

Let $H$ denote the ensemble hamiltonian which depends on the normalised probability density $p(x,t)$ and let $S(x,t)$ denote the canonically conjugate variable, 
\begin{equation}
H \equiv \int dx^{Nd} p ( h(p,S) + V), \label{ensHam}
\end{equation}
with $V$ some external potential. Hamilton's equations are
\begin{eqnarray}
{\partial p \over \partial t} &=& {\delta H \over \delta S} \; ,\label{ham1} \\
{\partial S \over \partial t} &=& -{\delta H \over \delta p} \; . \label{ham2} 
\end{eqnarray}
The Hamiltonian framework  was used earlier in \cite{HR1} but with a different set of axioms than those adopted in Ref.\cite{P5}. The reader is referred to \cite{HR1} for an extended discussion of the Poisson brackets and canonical transformations relevant also to the present context.

An explicit form for $h(p,S)$ can be constructed from the following reasonable  axioms: locality, separability, Galilean invariance, positivity of $H$, universality and simplicity \cite{P5}. Actually only rotational and translational invariance can be imposed on $h$ while invariance under boosts has to be imposed on the equations of motion.

Let me elaborate on the simplicity and universality conditions as they are less obvious. An example of what I mean by ``universality" is as follows: the normalisation of probability, $1 = \int dx^{Nd} p(x,t)$, implies that the dimensions of $p(x,t)$ depend on the dimensions of the configuration space. Thus $H$ can be universal only if $h$ is scale invariant, $h(\lambda p) =h(p)$. This scale invariance will ensure that the resulting equations of motion have a form independent of the number of particles. 

By ``simplicity" I mean that the hamiltonian should have  a minimum number of arbitrary constants, as befits fundamental dynamics. An example of this constraint is to let $h$ contain not more than two derivatives in any product of terms that appears in it. As each derivative involves an inverse length, this condition obviously restricts the number of new dimensional parameters, beyond the metric,  that can appear in the action. This specific condition will be referred to in brief as ``absence of higher number of derivatives" or ``AHD".

The criteria of universality and simplicity are what one expects of fundamental laws and they have been used, either explicitly or implicitly, at various times in physics: a classic example is Newton's law of universal gravitation to which reference will be made again at the end of the paper. 

Using the various axioms one obtains,  
\begin{equation}
h= g_{ij} \left( A (\partial_i S) (\partial_j S) + B{(\partial_i \log p) (\partial_j \log p)} \right) \,  \label{h3}
\end{equation}
with $A,B$ non-negative. Since $A$ can be absorbed in a redefinition of the metric, the final result  depends only on the single universal parameter $B$ with dimensions of action-squared. $B=0$ gives classical {\it ensemble} dynamics while $B>0$ leads, through Hamilton's equations,  to the quantum Hamilton-Jacobi equation (\ref{hj3}) and the continuity equation (\ref{cont}). The latter are equivalent to Schr\"odinger's equation with $B \sim \hbar^2$.

One of the most striking results of the above analysis is that the linearity of Schr\"odinger's equation is a consequence of the other axioms!

\section{Nonlinear Extensions, Gravity and Strings}
The physical axiomatic approach shows that the usual, non-relativistic and linear, Schr\"odinger's equation is a  deformation of classical ensemble  dynamics, the relevant parameter being $B$ in (\ref{h3}), but that both theories follow from the same set of axioms. 

If the simplicity axiom is relaxed to allow for higher number of derivatives, then dimensional analysis shows that  additional length scales  must be introduced. Some of the higher-derivative terms might be corrections to the {\it linear} Schr\"odinger equation such as what one would expect from the non-relativistic limit of the Klien-Gordon equation; the relevant length scale for these terms would be the Compton wavelength. However other allowable terms lead to {\it nonlinear} corrections to the Schr\"odinger equation and for those the length scale cannot be the Compton wavelength as that would conflict with empirical data \cite{P2}. Since we wish to keep universality, this implies that a higher-derivative extension of the non-relativistic Schr\"odinger equation, while keeping the other axioms, gives rise  to nonlinear corrections involving a new universal length scale $L$.

Nonlinear Schr\"odinger equations have been studied for many years with a period of high activity following Weinberg's contribution \cite{nonlin,nonlin2}. Various experiments indicated that the nonlinearities, if they existed at all, must be small \cite{Qexpts}. Some theoretical arguments initially suggested that the nonlinearities would result in pathologies but those conclusions were shown later 
by others to be premature \cite{prem}. Indeed, through a nonlinear gauge transformation one may convert the linear Schr\"odinger equation into an equivalent nonlinear form \cite{nonlin2}, so general arguments about the inconsistency of all nonlinear Schr\"odinger equations cannot be true (see also, for example, the discussion in \cite{alt} for 
 different perspectives and relevant references). Thus the linearity of Schr\"odinger's equation is still worth examining and, as will be apparent below, the new and promising experimental regime that should be explored is at high-energies or short distances. There is a vast literature on nonlinear Schr\"odinger equations constructed using different frameworks from  those adopted here, and studied with diverse motivations; the reader is referred to \cite{nonlin,nonlin2, S2}, and references therein, for a list of some of the papers.

Returning to the argument introducing a universal length scale, if one remains within the currently known forces of nature, then gravity enables a universal length scale to be formed, the Planck length, so possibly $L$ could be identified with that \cite{P3}. These arguments therefore suggest a link between gravity and the potential nonlinearity of Schr\"odinger's equation at short distances: of the {\it known} forces only gravity couples to all matter and so is the only candidate to form the {\it universal} length scale. However note that $L$ might turn out to be a larger length somewhere closer to a possible GUT theory \cite{P2}. 

It is interesting to note that some gravity induced corrections to Schr\"odinger's equation have been derived in the literature starting from the Wheeler-DeWitt equation \cite{GWD}, while others have introduced gravity-motivated corrections to Schr\"odinger's equation, see for example \cite{NS} and references therein. A nonlinear Schr\"odinger equation with a dissipative term \cite{db} has even been used to describe some low-energy dynamics of non-critical bosonic strings \cite{nav1}. 

If one wishes that, while relaxing the simplicity axiom,  all the possible nonlinear terms with higher number of derivatives be controlled by a {\it single} free parameter then the infinite number of terms must be summable into some compact expression. This would require one to make certain additional assumptions and to also relinquish  one or more of the above axioms. If one uses information theory motivations, which argue for ``unbiased" constructions, then a Schr\"odinger equation with a non-polynomial nonlinear correction, that is equivalent to a infinite number of derivatives, can be constructed \cite{P2}, but the equation breaks rotational symmetry. (Another reason for desiring a summation is that higher derivative terms that are scale-invariant might lead to singularities in the equations of motion at points where $p(x,t)$ vanishes).

Although a similar constructive approach for relativistic particle systems is not yet available, it is instructive to probe that domain using some  heuristic arguments. In the relativistic regime ``$c$", the speed of light in vacuum,   
appears now as a universal parameter but it is, as we know,  compatible with a linear quantum theory; its role is to covariantise the equations by transforming the dimensions of time to those of space. Thus if we imagine the nonlinear Schr\"odinger equation to be the nonrelativistic limit of a Dirac equation with a small nonlinearity, then the nonlinearity in the Dirac equation should involve higher-derivatives controlled by the length parameter $L$. Although such a nonlinear Dirac equation could still be formally Lorentz covariant, however generically it will no longer support the usual plane wave solutions with dispersion relation $E^2 = p^2 + m^2$. Thus one anticipates that at the relativistic level, potential nonlinear generalisations of quantum mechanics will lead to deformed dispersion relations for ``free" particles and thus practically to Lorentz symmetry breaking. Furthermore, keeping covariance and higher derivatives would imply higher-order time derivatives, which are typically problematic; but if only the spatial derivatives are of higher order then covariance is broken.  

The link between quantum nonlinearity and broken Lorentz symmetry has been discussed previously in \cite{P2,P3} using information theory ideas. The main appeal of all these arguments is that they identify Lorentz invariance as, possibly, the protecting symmetry for linear quantum theory, for then the scale of potential nonlinearities of quantum theory would be naturally associated with equally small potential violations of Lorentz invariance at short distances. For some other discussions in the literature concerning (non)linear quantum theory and short-distance physics,  see \cite{P2, S1} and references therein.

It is  noted that in the abovementioned study of non-critical bosonics strings where an effective nonlinear Schr\"odinger equation emerged,  the Lorentz invariance of the system was also spontaneously broken \cite{nav2}. For more references on nonlinear Schr\"odinger equations the interested reader is referred to \cite{nonlin,S2,S1}, and references therein.

\section{Summary and Outlook}

The ultimate aim of the various re-derivations and re-formulations of quantum theory differs among the authors who undertake those studies.  The investigations in \cite{P5,P1,P2,P3} were ventured primarily to understand the structure of Schr\"odinger's equation, but they also enable one to propose well-informed deformations of the standard theory that could be tested, or used for an eventual unification with spacetime. 
 
It has been shown that one may derive and understand the structure of Schr\"odinger's equation using standard classical concepts. The derivation summarised here has not assumed any classical action as a starting point, unlike  \cite{HR1,P1}. A similar construction  should be possible also for bosonic fields, probably using axioms as above to re-do the analysis in \cite{HR1}. Some refinement of the axioms would be necessary to include gravity, for example. Thus instead of the positivity axiom, one could demand instead that the continuity equation (\ref{cont}) is of the standard form; one must also be liberal in dropping total derivatives inside the integral for $H$.

It might also be useful to similarly study quantum mechanics in phase space, and even string/M theory, as this might give additional insight into those structures. Other applications of the methodology suggest themselves, for example in kinetic theory and hydrodynamics.

One may connect the discussion here to that in \cite{Reg,P1,P2} by interpreting the ensemble Hamiltonian $H$ as a generalised information (inverse uncertainty) measure: For fixed $S$, the ensemble Hamiltonian reaches a minimum when $p$ is uniform (maximum uncertainty). 

The physical axiomatic approach has suggested that minimal extensions of Schr\"odinger's equation, {\it within the discussed framework}, imply nonlinear corrections whose size is determined by a universal length scale. 
In this way one may systematically construct nonlinear quantum evolution equations, which are slight deformations of the linear theory, to probe potential new physics at short distances. 
The reason such nonlinear equations  might be useful for probing {\it small} deviations from current physics is because they would still obey many of the other physically motivated axioms that were used in their construction, and which are useful for interpreting the results. Some preliminary suggestions on the high-energy phenomenological consequences of a nonlinear quantum theory were made in \cite{P2,P3}. 

Various arguments  suggest that the nonlinear corrections are likely to be associated with a breaking of Lorentz symmetry, and the length scale possibly associated with gravity. Since there is currently significant interest in searches for Lorentz violation, some reviews are in  \cite{LV}, the potential nonlinear aspect of quantum theory in that regime should be kept in mind when interpreting the data. 

At a much more fundamental level however,  a better description of the ultra-microscopic world might involve something very different from a deformed Schr\"odinger-type equation.  A historical example of such a transition involves the  theory of gravity: While a small correction term can be added to Newton's inverse square law to account for the precession of planetary orbits, a reformulation by Einstein showed that gravity was more usefully viewed in an altogether different language; thus what appears simple today might become simpler tomorrow. Some recent suggestions for a sub-quantum theory may be found in \cite{subquantum1, subquantum2,smolinbook} and references therein.

\begin{theacknowledgments}
  I thank Prof. Andrei Khrennikov, and all the other organisers, for the opportunity to communicate this work at a stimulating workshop and for their hospitality. I also thank Harvendra Singh for helpful discussions  on string theory. \end{theacknowledgments}

\end{document}